\documentclass[iop]{emulateapj}

\def\ltsima{$\; \buildrel < \over \sim \;$}
\def\simlt{\lower.5ex\hbox{\ltsima}}
\def\gtsima{$\; \buildrel > \over \sim \;$}
\def\simgt{\lower.5ex\hbox{\gtsima}}
\def\kms{{\rm\,km\,s^{-1}}}

\def\kpc{{\rm\,kpc}}

\def\pc{{\rm\,pc}}

\def\deg{^\circ}
\def\s{\ifmmode \widetilde \else \~\fi}
\def\={\overline}

\def\spose#1{\hbox to 0pt{#1\hss}}

\def\lta{\mathrel{\spose{\lower 3pt\hbox{$\mathchar"218$}}
     \raise 2.0pt\hbox{$\mathchar"13C$}}}
\def\gta{\mathrel{\spose{\lower 3pt\hbox{$\mathchar"218$}}
     \raise 2.0pt\hbox{$\mathchar"13E$}}}
\def\Dt{\spose{\raise 1.5ex\hbox{\hskip3pt$\mathchar"201$}}}    % upper case
\def\dt{\spose{\raise 1.0ex\hbox{\hskip2pt$\mathchar"201$}}}    % lower case

\def\dotsfill{\leaders\hbox to 1em{\hss.\hss}\hfill}

\def\Gyr{{\rm\,Gyr}}
\def\FeH{{\rm[Fe/H]}}

\shorttitle{A new Milky Way satellite}
\shortauthors{B. P. M. Laevens et al.}

\begin{document}


\title{A new faint Milky Way satellite discovered in the Pan-STARRS1 $3\pi$ survey}


\author{Benjamin P. M. Laevens$^{1,2}$, Nicolas F. Martin$^{1,2}$, Rodrigo A. Ibata$^{1}$, Hans-Walter Rix$^2$, Edouard J. Bernard$^3$, Eric F. Bell$^{4}$, Branimir Sesar$^2$, Annette M. N. Ferguson$^3$, Edward F. Schlafly$^2$, Colin T. Slater$^4$, William S. Burgett$^5$, Kenneth C. Chambers$^6$, Heather Flewelling$^6$, Klaus A. Hodapp$^6$, Nicholas Kaiser$^6$, Rolf-Peter Kudritzki$^6$, Robert H. Lupton$^7$,  Eugene A. Magnier$^6$, Nigel Metcalfe$^8$, Jeffrey S. Morgan$^6$, Paul A. Price$^7$, John L. Tonry$^6$, Richard J. Wainscoat$^6$, Christopher Waters$^6$}
%Branimir Sesar$^2$,  Eric F. Bell$^3$, Colin T. Slater$^3$, Edward F. Schlafly$^2$ et al.}
%Edouard J. Bernard$^3$, Colin T. Slater$^4$, Eric F. Bell$^4$, Edward F. Schlafly$^2$,  Annette M. N. Ferguson$^3$, William S. Burgett$^5$, Kenneth C. Chambers$^5$, Larry Denneau$^5$, Peter W. Draper$^6$, Nicholas Kaiser$^5$,  Rolf-Peter Kudritzki$^5$, Eugene A. Magnier$^5$, Nigel Metcalfe$^6$, Jeffrey S. Morgan$^5$, Paul A. Price$^7$, William E. Sweeney$^5$, John L. Tonry$^5$,  Richard J. Wainscoat$^5$, Christopher Waters$^5$}

%Paolo Bianchini$^2$, Douglas P. Finkbeiner$^5$, Edouard J. Bernard$^4$, William S. Burgett$^6$, Kenneth C. Chambers$^6$, Klaus W. Hodapp$^6$, Nicholas Kaiser$^6$, Rolf-Peter Kudritzki$^6$, Eugene A. Magnier$^6$, Jeffrey S. Morgan$^6$, Paul A. Price$^7$, John L. Tonry$^6$, Richard J. Wainscoat$^6$}
\email{benjamin.laevens@astro.unistra.fr}

\altaffiltext{1}{Observatoire astronomique de Strasbourg, Universit\'e de Strasbourg, CNRS, UMR 7550, 11 rue de l'Universit\'e, F-67000 Strasbourg, France}
\altaffiltext{2}{Max-Planck-Institut f\"ur Astronomie, K\"onigstuhl 17, D-69117 Heidelberg, Germany}
\altaffiltext{3}{Institute for Astronomy, University of Edinburgh, Royal Observatory, Blackford Hill, Edinburgh EH9 3HJ, UK}
\altaffiltext{4}{Department of Astronomy, University of Michigan, 500 Church St., Ann Arbor, MI 48109, USA}
\altaffiltext{5}{GMTO Corporation, 251 S. Lake Ave, Suite 300, Pasadena, CA  91101, USA}
\altaffiltext{6}{Institute for Astronomy, University of Hawaii at Manoa, Honolulu, HI 96822, USA}
\altaffiltext{7}{Department of Astrophysical Sciences, Princeton University, Princeton, NJ 08544, USA}
\altaffiltext{8}{Department of Physics, Durham University, South Road, Durham DH1 3LE, UK}
%\altaffiltext{4}{Department of Physics and Astronomy, University of California, Los Angeles, CA 90095-1547, USA}
%\altaffiltext{5}{Harvard-Smithsonian Center for Astrophysics, 60 Garden Street, Cambridge, MA 02138, USA}

\begin{abstract}
We present the discovery of a faint Milky Way satellite, Laevens~2/Triangulum~II, found in the Panoramic Survey Telescope And Rapid Response System (Pan-STARRS 1) 3 $\pi$ imaging data and confirmed with follow-up wide-field photometry from the Large Binocular Cameras. The stellar system, with an absolute magnitude of $M_V=-1.8\pm0.5$, a heliocentric distance of $30^{+2}_{-2}\kpc$, and a half-mass radius of $34^{+9}_{-8}\pc$, shows remarkable similarity to faint, nearby, small satellites such as Willman 1, Segue 1, Segue 2, and Bo\"otes II. The discovery of Laevens~2/Triangulum~II further populates the region of parameter space for which the boundary between dwarf galaxies and globular clusters becomes tenuous. Follow-up spectroscopy will ultimately determine the nature of this new satellite, whose spatial location hints at a possible connection with the complex Triangulum-Andromeda stellar structures.
\end{abstract}


\section{Introduction}
The last couple of decades saw the discovery of numerous satellites in the Milky Way (MW) halo. While the Sloan Digital Sky Survey (SDSS \citep{york00}) satellite discoveries have provided us with greater observational constraints in our backyard, especially to understand the faint end of galaxy formation in the preferred cosmological paradigm of $\Lambda$CDM \citep{belokurov13}, they have also led to debates about the nature of the faintest satellites \citep{gilmore07}. It has become apparent that the previously clear distinction between the compact globular clusters (GCs) and the brighter, more extended, and dark-matter dominated dwarf galaxies (DGs), blurs out for faint systems \citep{willman12}. This is exemplified by the discoveries of Willman~1 \citep[Wil1;][]{willman05a} and Segue~1 \citep[Seg1;][]{belokurov07a}, followed up by those of Bo\"otes~II \citep[BooII;][]{walsh08}, and Segue~2 \citep[Seg2;][]{belokurov09}, all nearby satellites within 25--45 \kpc, and just slightly larger than extended outer halo GCs. At the same time, these systems are fainter than most GCs and all the other DGs. Theoretical expectations show that these objects could well be the faintest DGs and that tens or hundreds of DGs with these properties could populate the Milky Way halo \citep{tollerud08,hargis14}. As of yet, just two objects have been found in the PS1 survey \citep{laevens14}, reinforcing the tension between theory and observations \citep{klypin99}.Only the closest DGs would be detected with current photometric surveys \citep{koposov07, walsh09}. Spectroscopic studies do show that the faint systems found so far are dynamically hotter than their mere stellar content would imply, hinting that they are indeed DGs \citep{martin07a, willman11, simon11, kirby13}. However, the low velocity dispersion of these satellites ($<4\kms$), combined with the possibly large impact of binaries \citep{mcconnachie10}, the complexity of disentangling member stars from foreground contaminants, and the overall dimness of their member stars renders any definite conclusion difficult.

\begin{figure*}
\begin{center}
\includegraphics[width=0.52\hsize,angle=270]{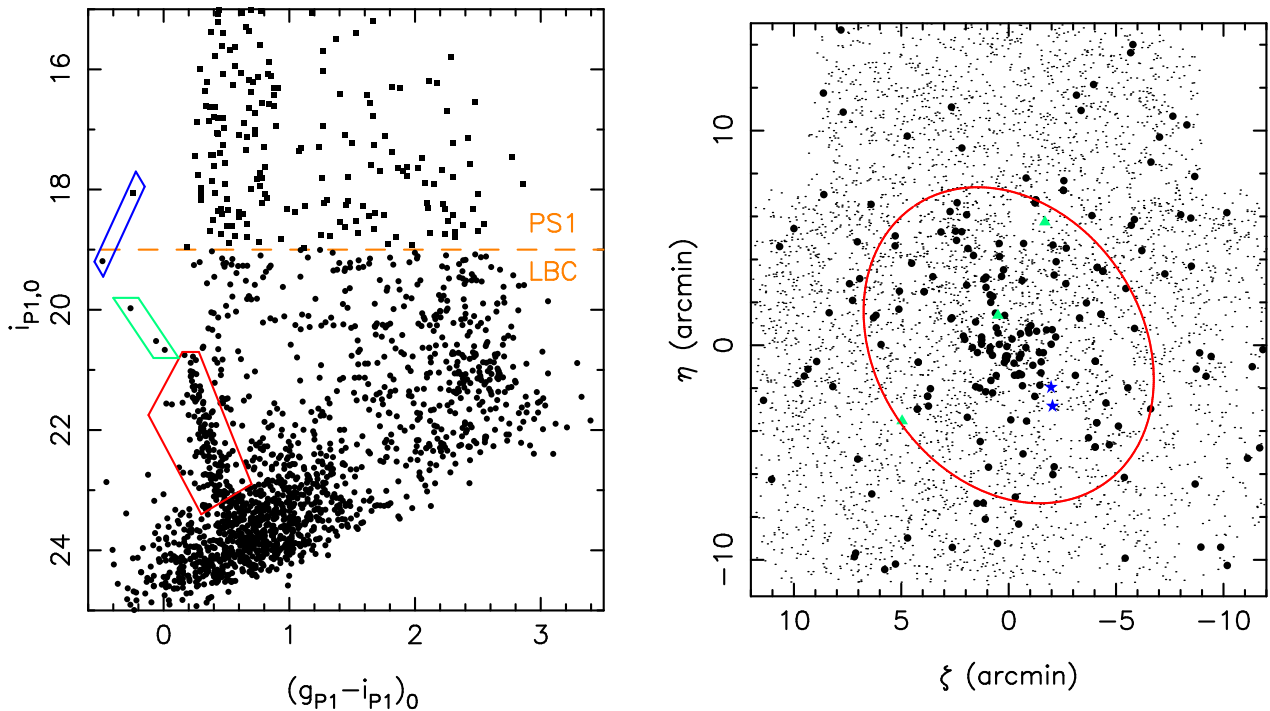}
\caption{\label{CMDs}: \emph{Left}: The combined PS1-LBC CMD of all sources within the central $2r_h$ region of Lae~2/Tri~II. The single epoch PS1 photometry was used at the bright end ($i_{\mathrm{P1},0}<19$, squares), with LBC photometry supplementing the faint end $i_{\mathrm{P1},0}>19$ (large dots). The orange dashed line indicates the separation between the LBC and PS1 data. The red box highlights the clear main sequence of the stellar system, the blue box indicates two possible HB stars and the green box identifies likely blue stragglers. \emph{Right}: Spatial distribution of all sources corresponding to the CMD on the left. Large dots correspond to the stars falling within the red CMD box in the left panel and show a clear overdensity. The two blue stars indicate the possible HB stars whereas the red ellipse corresponds to the region within the favored two half-mass radius of the system, as inferred below.}
\end{center}
\end{figure*}

Here, we present the discovery of another faint MW satellite, Laevens~2/Triangulum~II  \footnote{In the absence of spectroscopic confirmation, we wish to remain agnostic about the nature of this object and therefore propose a double name. For future reference in this paper, we abbreviate to Lae~2/Tri~II.} 
%The definitive name of this object may be decided when the spectroscopic follow-up proves conclusive. For future reference in this paper, we abbreviate to Lae~2/Tri~II.}
, with very similar photometric properties to Wil1, Seg1, BooII, and Seg2. The new system was found in our ongoing effort to mine the Pan-STARRS 1 (PS1) $3\pi$ survey for localized stellar overdensities. This letter is structured as follows: in section~2, we describe the PS1 survey along with the detection method that led to the discovery. We continue by discussing follow-up imaging obtained with the Large Binocular Cameras in section~3. We discuss the nature of the satellite and its implication in section~4. In the final section, we summarize and conclude our results.

In this paper, magnitudes are dereddened using the \citet{schlegel98} maps, adopting the extinction coefficients of \citet{schlafly11}. A heliocentric distance of $8\kpc$ to the Galactic center is assumed.

\section{The $3\pi$ PS1 Survey and discovery}
With a spatial extent encompassing three quarters of the sky ($\delta>-30\deg$), PS1 (K. Chambers et al., in preparation) gives us an unprecedented panoptic view of the MW and its surroundings. Over the course of 3.5 years, the 1.8m telescope, equipped with its 1.4-gigapixel camera covering a 3.3-degree field of view, has collected up to four exposures per year in each of 5 bands ($g_\mathrm{P1}r_\mathrm{P1}i_\mathrm{P1}z_\mathrm{P1}y_\mathrm{P1}$; \citealt{tonry12}). A photometric catalogue is automatically generated with the Image Processing Pipeline \citep{magnier06,magnier07,magnier08}, once the individual frames have been downloaded from the summit. The preliminary stacked photometry used in this paper has a $g_\mathrm{P1}$ depth (23.0) that is comparable to SDSS $g$-band depth and $r_\mathrm{P1}$/$i_\mathrm{P1}$ observations that reach $\sim0.5$/$\sim1.0$~magnitude fainter: 22.8, 22.5 for r and i respectively \citep{metcalfe13}.

Inspired by past searches for small stellar overdensities in MW and M31 surveys, we apply a convolution technique (Laevens et al., in preparation), successfully used to find new GCs and DGs in the SDSS \citep{koposov07,walsh09}. In a nutshell, we build a mask in $(r-i,i)$ color-magnitude space to isolate potential metal-poor, old, and blue member stars that could belong to a MW satellite at a chosen distance. This mask is applied to star-like sources in the stacked PS1 photometric catalog. We then convolve the distribution of isolated sources with two Gaussian spatial filters: a positive Gaussian tailored to the size of the overdensities we are searching for ($2'$, $4'$, or $8'$) and a negative Gaussian with a much larger kernel ($14'$, $28'$, or $56'$), to account for the slowly-varying contamination of sources that fall within the color-magnitude mask. By convolving the data with the sum of these two (positive and negative) filters and accounting for the survey's spatial incompleteness on the arcminute scale, we obtain maps tracking stellar over- and under-densities in PS1. We convert these density maps into maps of statistical significance by comparison with the neighboring regions after cycling through distances and filter sizes. This procedure already led to the discovery of Laevens~1\footnote{Following the naming  convention established in \citet{bianchini15}} \citep{laevens14} also discovered concomitantly as Crater within the ATLAS survey by \citet{belokurov14b}. The new satellite, Lae~2/Tri~II, is located $\sim20\deg$ East of M31 and appears as a $5.2\sigma$ detection, only slightly higher than our significance criteria of $5\sigma$\footnote{These also include a check that potential detections do not also correspond to a significant overdensity of background galaxies \citep{koposov07}.} tailored to weed out spurious detections.

\section{Follow-Up}
\begin{figure*}
\begin{center}
\includegraphics[width=0.52\hsize,angle=270]{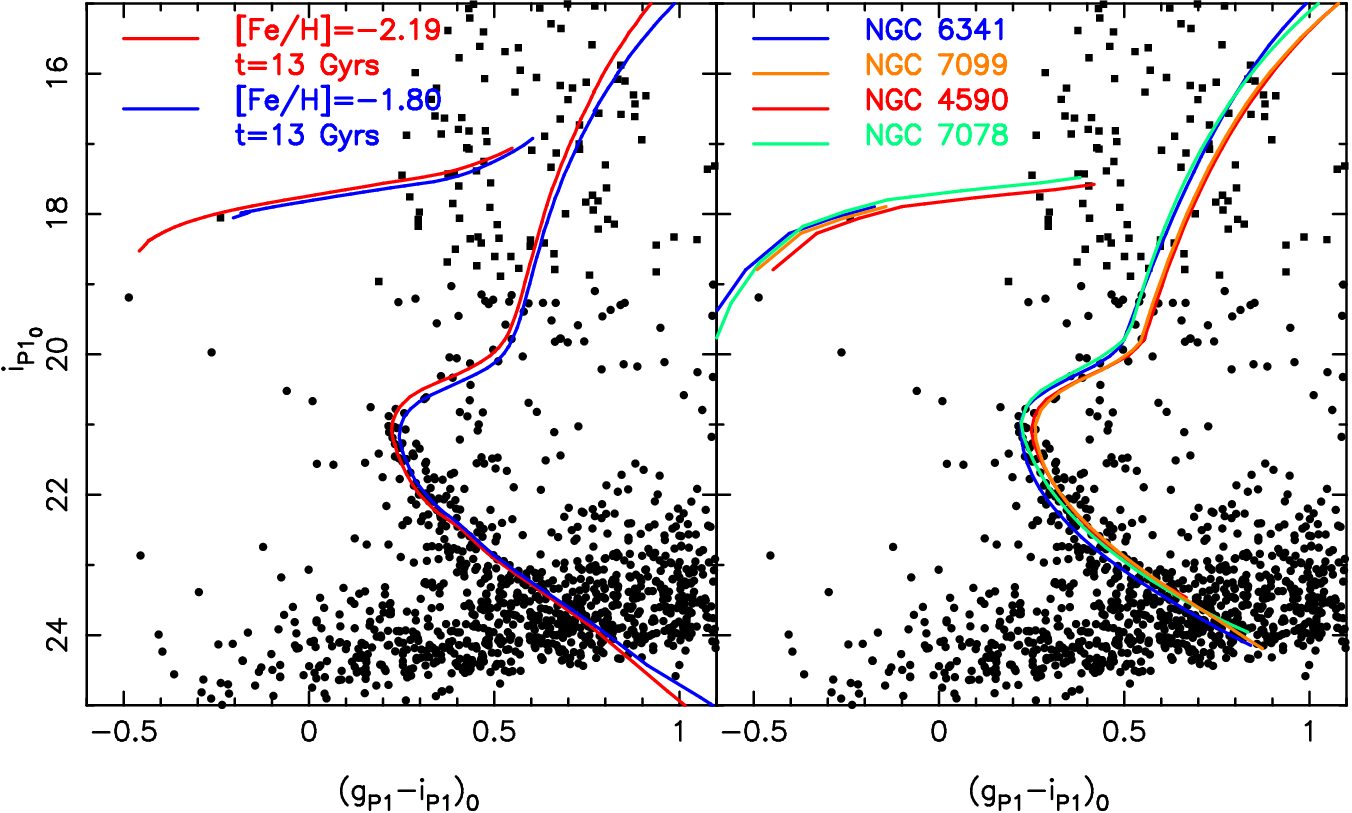}
\caption{\label{iso}\emph{Left}: The CMD of Lae~2/Tri~II within $2r_h$, with best-fit isochrones overplotted. Both isochrones have an age of 13 Gyr and are shifted to a distance modulus of 17.3. The red/blue isochrones have metallicities $\FeH=-2.19$ and $\FeH=-1.80$ , respectively. \emph{Right}: The same CMD of Lae~2/Tri~II with four metal-poor fiducial isochrones from GCs observed in PS1 \citep{bernard14}, dereddened assuming the reddening values of \citep{harris10} and shifted to match the observed features. The green fiducial (NGC 7078; $\FeH=-2.37$), shifted to a distance-modulus of 17.5 best represents the features of the new stellar system, with the other fiducials i.e. red: NGC 4590 ($\FeH=-2.23$), blue: NGC 6341 ($\FeH=-2.31$), orange: NGC 7099 ($\FeH=-2.27$) appearing too red or to blue to accurately reproduce the MS and MSTO. NGC 4590 is shifted to a distance-modulus of 17.5, whereas NGC 6341 and 7099 are at 17.3.
}
\end{center}
\end{figure*}

To confirm the nature and the properties of Lae~2/Tri~II, follow-up imaging was obtained with the Large Binocular Camera (LBC) on the Large Binocular Telecope (LBT), located on Mount Graham, USA during the night of October 17--18 2014. With its $23'\times25'$ field of view and equipped with 4 CCDs, the LBC are ideal to follow-up MW satellites that usually span a few arcminutes on the sky. Imaging was conducted in the $g$ and $i$ bands, making use of the time-saving dual (binocular) mode using the red and blue eye simultaneously. Six dithered 200s sub-exposures were acquired in each band, with a seeing of $~1''$. The field was centered on the location of Lae~2/Tri~II.

The images were processed and the photometry performed using a version of the CASU pipeline \citep{irwin01} updated to work on LBC data. The instrumental magnitudes were calibrated onto the PS1 system ($g_\mathrm{P1}$ and $i_\mathrm{P1}$), by comparison with the PS1 single epoch data \citep{schlafly12} to derive the relevant color equations. The final LBC photometry reaches more than 2~magnitudes deeper than the stacked PS1 data. The left-hand panel of Figure~\ref{CMDs} shows the combined PS1/LBC color-magnitude diagram (CMD) of all stars within 2 half-light radii ($\pm2 r_h$), for which all sources brighter than $i_{\mathrm{P1},0} =19.0$ are taken from the PS1 single epoch photometry so as to extend the CMD beyond the saturation limit of the LBC photometry. The main sequence (MS) of an old and metal-poor stellar system is readily visible, with a clear turn off at $i_{\mathrm{P1},0} \sim 20.8$. Given the low density of the MS, the red giant branch (RGB) of the stellar system is very likely sparsely populated and hidden within the foreground contamination of bright stars. However, two blue stars are consistent with being blue horizontal branch (HB) stars and are highlighted in the CMD. The other panel of Figure~\ref{CMDs} presents the spatial distribution of sources in the four chips, with a red ellipse indicating the satellite's two half-light radii extent (as determined by the structural parameter analysis, see section 4). Stars with colors and magnitudes consistent with the MS are shown as large dots  and reveal a clear spatial overdensity. Likely blue straggler stars are identified and given by the green triangles. The two blue stars correspond to the potential HB stars; their location close to the center of the stellar overdensity supports them being member HB stars.

\begin{figure*}
\begin{center}
\includegraphics[width=0.42\hsize,angle=270]{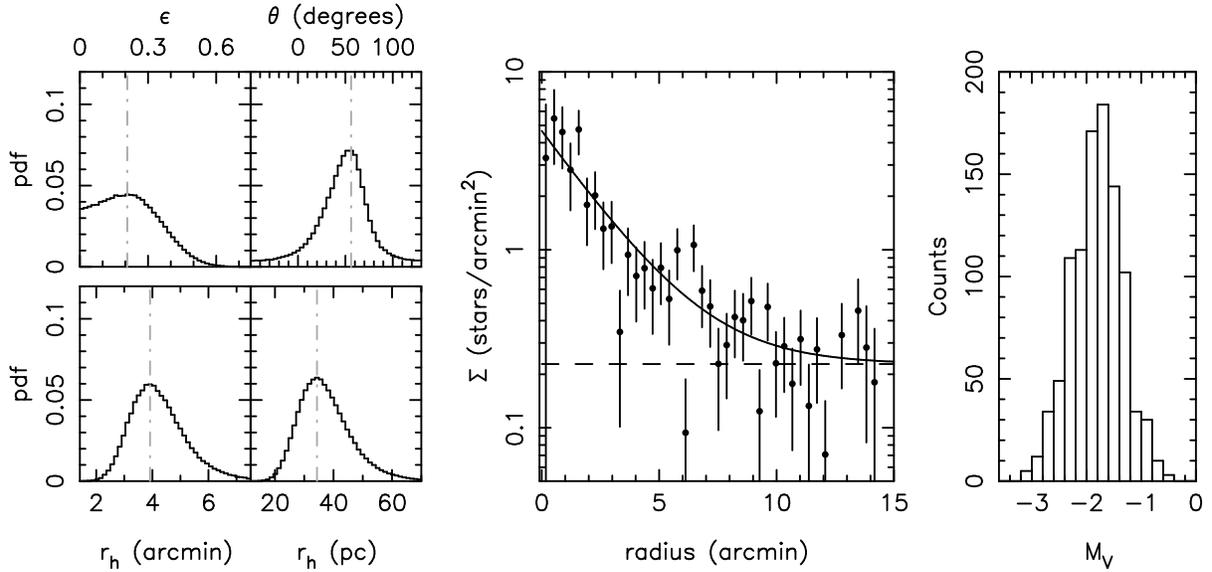}
\caption{\label{profile} \emph{Left}: Probability distribution functions for the ellipticity ($\epsilon$), the position angle ($\theta$), the angular and the physical half-mass radii ($r_h$) of Lae~2/Tri~II (from top left to bottom right). \emph{Middle}: Comparison between the favored radial distribution profile (full line) and the data, binned according to the preferred structural parameters (dots). The error bars assume Poissonian uncertainties and the dashed line represents the field density. \emph{Right}: Probability distribution function of the absolute magnitude of Lae~2/Tri~II in the $V$ band ($M_V=-1.8\pm0.5$).}
\end{center}
\end{figure*}

\section{Properties of the stellar system}
Since the RGB and HB are so sparsely populated, an investigation into the presence of member RR Lyrae stars in the multi-epoch PS1 data unsurprisingly led to no candidate from which to derive a distance estimate. However, due to the well defined MS and MS turnoff (MSTO) at $i_{\mathrm{P1},0} \sim 20.8$, a reliable distance estimate can nevertheless be determined through a comparison with isochrones and fiducials by eye (Figure~\ref{iso}). We assess the stellar system's metallicity, age, and distance modulus by first cycling through \textsc{Parsec} isochrones \citep{bressan12} for a metallicity range $-2.2<\FeH<-1.3$ ($Z = 0.0001$ to $0.0007$, assuming $Z_{\odot}=0.152$) and an age between 9 and $13\Gyr$. We investigate the effects that various metallicities and ages have on the distance determination, by cycling through distance-modulus steps of 0.1 between 16.9 and 17.5. The isochrone which best represents the CMD features is a metal-poor, old isochrone ($\FeH=-2.19$ and age of 13 Gyrs), for a distance modulus of 17.3. We further strengthen these conclusions by comparing the CMD of Lae~2/Tri~II with the fiducials from 13 GCs and 3 Open Clusters of varying metallicity, derived directly from the PS1 data \citep{bernard14}. The 4 most metal-poor GCs of the sample provide a good fit to the MS and MSTO of Lae~2/Tri~II provided they are shifted to distance moduli in the 17.3-17.5 range.
%Its metallicity ($\FeH=-2.37$) is compatible for that favored by the \textsc{Parsec} isochrones.

Combining these two analyses by averaging the six best-fits fiducials and isochrones, we therefore conclude that Lae~2/Tri~II is old and metal-poor, and is located at a distance modulus of $17.4^{+0.1}_{-0.1}$, which translates to a heliocentric distance of  $30^{+2}_{-2}\kpc$, or $36^{+2}_{-2}\kpc$ from the Galactic center. In both cases (isochrone and fiducial), the HB of the favored track also overlaps almost exactly with the two potential HB stars. Please note that the uncertainty in the distance measurement is propagated through for the derivation of the structural parameters, further detailed in the next few paragraphs. We also draw to the reader's attention that a more involved analysis of `CMD-fitting' would likely enhance the quality of the distance measurement; however, the limitation of the field of view prevent us from obtaining a large enough sample of background stars to perform such an analysis.

\begin{table}
\caption{\label{properties}Properties of Lae~2/Tri~II}
\begin{center}
\begin{tabular}{cc}
\hline
$\alpha$ (J2000) & 02:13:17.4\\
$\delta$ (J2000) & +36:10:42.4\\
$\ell$ & $141.4\deg$\\
$b$ & $-23.4\deg$\\
Distance Modulus & $\sim17.4^{+0.1}_{-0.1}$\\
%$17.3\pm0.15$\\
Heliocentric Distance & $30^{+2}_{-2}\kpc$\\
%$32\pm2\kpc$\\
Galactocentric Distance & $36^{+2}_{-2}\kpc$\\
%42 \pm0\kpc$ \\
$M_{V}$ & $-1.8\pm0.5$\\
\FeH & $\sim-2.2$\\
Age & $\sim13\Gyr$\\
$E(B-V)^{a}$ & 0.081\\
 %& Exponential profile\\
Ellipticity & $0.21^{+0.17}_{-0.21}$\\
Position angle (from N to E) & $56^{+16}_{-24}$$\deg$\\
$r_{h}$ & $3.9^{+1.1}_{-0.9}$$'$\\
$r_{h}$ & $34^{+9}_{-8}\pc$\\
\hline
$^a$ from \citet{schlegel98} &\\
\end{tabular}
\end{center}
\end{table}

We derive the structural parameters of Lae~2/Tri~II by using a modified version of the technique described in \citet{martin08b}, updated in Martin et al. (2015, submitted). The updated technique allows for a Markov Chain Monte Carlo (MCMC) approach, based on the likelihood of a family of exponential radial density profiles (allowing for flattening and a constant contamination over the field) to reproduce the distribution of the system's MS stars. The parameters of the model are: the centroid of the system, its ellipticity\footnote{The ellipticity is here defined as $1-b/a$ with $a$ and $b$ the major and minor axis scale lengths, respectively.}, its position angle (defined as the angle of the major axis from North to East), its half-mass radius\footnote{Note that, assuming no mass segragation in the system, the half-mass radius is equivalent to the half-light radius.}, $r_h$, and the number of stars, $N^*$, within the chosen CMD selection box. Although it is located in the Triangulum constellation, Lae~2/Tri~II is so far from both M31 ($\sim$20 degrees) and M33 ($\sim$10 degrees) that any contamination by M31 or M33 stellar populations is vanishingly small and does not impact our results. The resulting probability distribution functions are presented in the left-most panels of Figure~\ref{profile} for the most important parameters and summarized in Table~1. Figure ~\ref{profile} also shows a favorable comparison of the preferred exponential profile with the data binned following the preferred structural parameters. 

\begin{figure*}
\begin{center}
\includegraphics[width=0.5\hsize,angle=270]{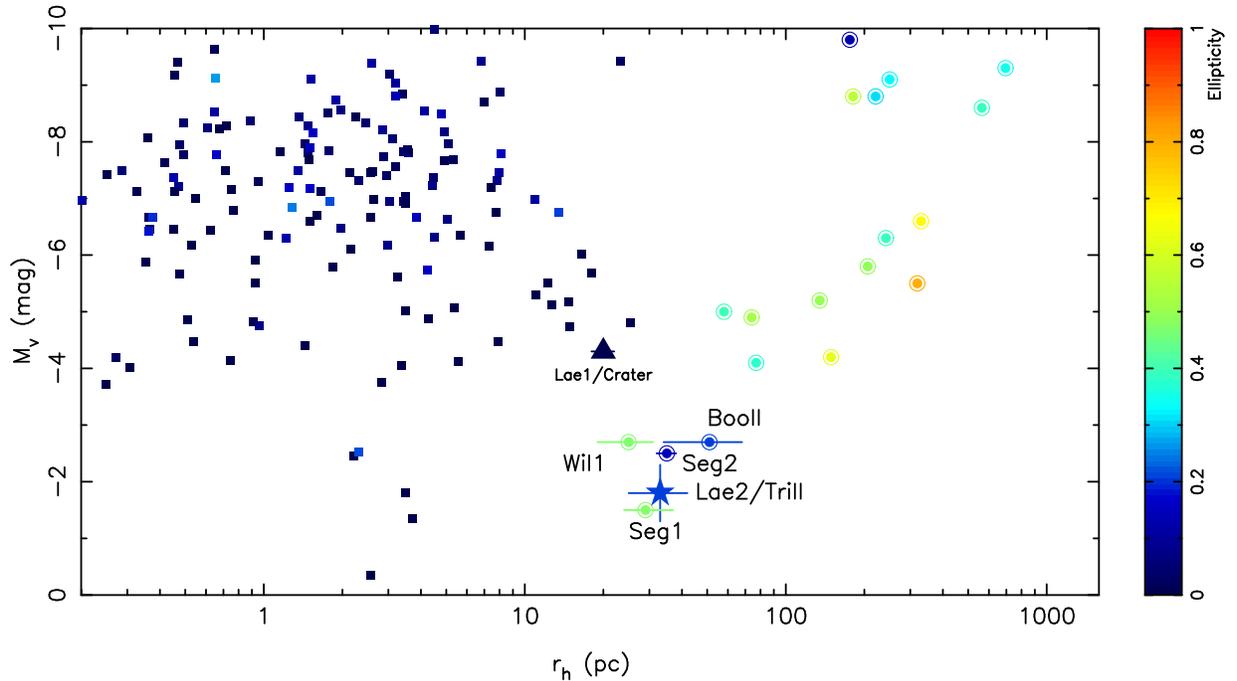}
\caption{\label{GCs_DGs_Props}The distribution of MW satellites in size--magnitude space. GCs are shown as squares, DGs are shown as circles and Lae~2/Tri~II is represented by the large star symbol. The color scale indicates the ellipticity of the various satellites. Lae~2/Tri~II's ellipticity and half-mass radius show very similar values to those of the four satellites: Seg1, Seg2, Bo\"oII and Wil1. Finally, we also indicate the recently discovered MW satellite, Laevens~1/Crater given by the large triangle. The data for the GCs were taken from \citet{harris10} and the DGs from \citet{mcconnachie12}.}
\end{center}
\end{figure*}

It should however be noted that the properties of Lae~2/Tri~II as observed by the LBC could be slightly biased by the low contrast of the stellar overdensity. Indeed, \citet{munoz12} show that satellite properties are most accurately measured when the central density of stars relative to that of the background is larger than 20, which is not the case here. Deeper data would be necessary to strengthen our size measurement.

%To determine the absolute magnitude of the stellar system, we follow the same procedure we used for Laevens~1/Crater \citet{laevens14} and was initially described in \citet{martin08b}. After drawing a value of $N^*$ from the structural parameter chain, we sample the CMD of the best \citet{bressan12} isochrone (see Figure~\ref{iso}), with its associated luminosity function and photometric uncertainties, until it contains $N^*$ stars in the CMD selection box used for the structural parameter analysis. Adding up the flux of all the stars drawn in this artificial yields the absolute $g_\mathrm{P1}$- and $i_\mathrm{P1}$-band magnitudes of Lae~2/Tri~II ($M_g = $ and $M_i = $, which converts to $M_V=-1.8\pm0.5$. This technique has the benefit of accounting for the effect of sampling such a small population of stars may have on the determination of the system's magnitude (i.e. CMD `shot-noise'; \citealt{martin08b}).

To determine the absolute magnitude of the stellar system, we follow the same procedure we used for Laevens~1/Crater \citep{laevens14} as was initially described in \citet{martin08b}. After drawing a value of $N^*$ from the structural parameter chain, we sample the CMD of the best-fitting \citet{bressan12} isochrone (see Figure~\ref{iso}), with its associated luminosity function and photometric uncertainties, until it contains $N^*$ stars in the CMD selection box used for the structural parameter analysis. Adding up the flux of all the stars drawn in this artificial CMD yields the absolute $g_\mathrm{P1}$- and $i_\mathrm{P1}$-band magnitudes of Lae~2/Tri~II ($M_g = -1.7\pm0.5$ and $M_i = -2.1\pm0.5$), which converts to $M_V=-1.8\pm0.5$. This technique has the benefit of accounting for the effect of sampling such a small population of stars may have on the determination of the system's magnitude (i.e. CMD `shot-noise'; \citealt{martin08b}).

\section{Discussion and Conclusion}

We have presented the discovery of a new MW satellite, Lae~2/Tri~II, discovered within the PS1 $3\pi$ data and confirmed from deep and wide LBC follow-up. Located at a heliocentric distance of $30^{+2}_{-2}\kpc$, this system is very faint ($M_V=-1.8\pm0.5$), old ($\sim13$ Gyrs), metal-poor ($\FeH\sim-2.2$), small ($34^{+9}_{-8}\pc$), and mildly elliptical ($0.21^{+0.17}_{-0.21}$). Figure~\ref{GCs_DGs_Props} places Lae~2/Tri~II in relation to other MW GCs and DGs. This new system's magnitude and half-mass radius are very similar to the properties of the faint satellites Seg1, Seg2, Wil1, and Bo\"oII, which were all recently discovered in the SDSS. Ultimately, high quality spectroscopic follow-up and an assessment of its dynamics are necessary to confirm the nature of this new satellite. However, its similarity in distance, size, absolute magnitude, age, and metallicity to those of Wil1, Seg1, Bo\"oII, and Seg2, that all have larger velocity dispersion than implied by their tiny stellar mass \citep{martin07a,simon11,willman11,kirby13} hints that Lae~2/Tri~II could well be another one of these systems that appear to populate the faint end of the galaxy realm.

It is also worth noting that the location of Lae~2/Tri~II, $(\ell,b) = (141.4\deg,-23.4\deg)$, $\sim20\deg$ East of M31, places it within the Triangulum-Andromeda stellar structure(s) \citep[TriAnd;][]{majewski04b,rocha-pinto04,sheffield14}. Although this MW halo stellar overdensity is very complex, with evidence for multiple substructures (\citealt{bonaca12,martinc13}; \citealt{martin14}), it spans a large enough distance range to encompass Lae~2/Tri~II ($\sim15-35\kpc$). A recent spectroscopic study of stars within TriAnd by \citet{deason14} confirmed that, as initially proposed by \citet{belokurov09}, Seg2 is also likely embedded within it and follows a systematic trend of these faint satellites being part of MW halo stellar streams. Seg1 has been proposed to be tied to the Orphan Stream \citep{newberg10}, though differences in abundance patterns between both have also been observed \citep{vargas13,casey14}. Similarly Bo\"oII's distance and radial velocity are compatible with it being part of the Sagittarius stream \citep{koch09}, whereas high resolution abundance measurements for Bo\"oII stars  question this association \citep{koch14}. It remains possible, however, that the small stellar systems were satellites of the larger, now disrupted progenitor of these stream, thereby alleviating the need for them to share similar abundances. In this context, it is particularly interesting that Lae~2/Tri~II is situated on the linear extrapolation of the Pan-Andromeda Archaeological Survey (PAndAS) MW stream \citep{martin14}, $10\deg$ beyond the PAndAS footprint where this dwarf galaxy remnant was discovered. The stream and satellite are not aligned; however, the uncertainties on the position angle and ellipticity are not conclusive in ruling this out. Here as well, spectroscopy is necessary to derive the systemic velocity of Lae~2/Tri~II and confirm it is compatible with the global motion of Triangulum-Andromeda and, in particular, with the velocity of the PAndAS MW stream. Follow-up will help reinforce or disprove such a hypothesis.

Although PS1 is only slightly deeper than the SDSS, the extra coverage provided by its $3\pi$ footprint leaves hope for more discoveries of faint objects like Lae~2/Tri~II. Building up the statistics of these systems through more discoveries in current (PS1) and (DES; \citealt{des05}) or future surveys (LSST; \citealt{tyson02}) is essential if we are to understand the true nature of these incredibly faint stellar systems that can only be found in the MW surroundings.

\acknowledgments
B.P.M.L. acknowledges funding through a 2012 Strasbourg IDEX (Initiative d'Excellence) grant, awarded by the French ministry of education. N.F.M. and B.P.M.L. gratefully acknowledges the CNRS for support through PICS project PICS06183. H.-W.R. and E.F.T acknowledge support by the DFG through the SFB 881 (A3). E.F.B. and C.T.S. acknowledge support from NSF grant AST 1008342. We thank the LBT and the observers for the splendid job that enabled the confirmation of this object.
%We thank the LBT for the follow-up that enabled the confirmation of this object as well as the LBT observers for doing such a splendid job.

The Pan-STARRS1 Surveys have been made possible through contributions of the Institute for Astronomy, the University of Hawaii, the Pan-STARRS Project Office, the Max-Planck Society and its participating institutes, the Max Planck Institute for Astronomy, Heidelberg and the Max Planck Institute for Extraterrestrial Physics, Garching, the Johns Hopkins University, Durham University, the University of Edinburgh, Queen's University Belfast, the Harvard-Smithsonian Center for Astrophysics, the Las Cumbres Observatory Global Telescope Network Incorporated, the National Central University of Taiwan, the Space Telescope Science Institute, the National Aeronautics and Space Administration under Grant No. NNX08AR22G issued through the Planetary Science Division of the NASA Science Mission Directorate, the National Science Foundation under Grant No. AST-1238877, the University of Maryland, and Eotvos Lorand University (ELTE).

%\bibliography{Biblio}

\bibliographystyle{apj}

% Bibtex will create a .bbs file in the directory and before sending to the editor, I should replace the bibliography call by this file.

\end{document}